\documentclass[aps,prb,amsmath,amssymb,showpacs,reprint,superscriptaddress]{revtex4-1}
\pdfoutput=1 %arXiv
 
\usepackage{color, graphicx}
\usepackage[version=3]{mhchem} % Formula subscripts using \ce{}
\usepackage{dcolumn}% Align table columns on decimal point
\usepackage{bm}% bold math
\usepackage[]{units} % u can enter units of measurement  \unit[3000]{\icm} 
\usepackage{hyperref}

\newcommand{\icm}{cm$^{-1}$}
\newcommand{\mapi}{\ce{MAPbI3}}
\newcommand{\grun}{Gr\"{u}neisen}

\begin{document}

\title{Lattice dynamics and vibrational spectra of the orthorhombic, tetragonal and cubic phases of methylammonium lead iodide}

\author{Federico Brivio}
\author{Jarvist M. Frost}
\author{Jonathan M. Skelton}
\author{Adam J. Jackson}
\author{Oliver J. Weber}
\author{Mark T. Weller}
\affiliation{Centre for Sustainable Chemical Technologies and Department of Chemistry, University of Bath, Claverton Down, Bath BA2 7AY, UK}

%\author{Mariano Campoy-Quiles}
%\affiliation{Institut de Ci\`encia de Materials de Barcelona (ICMAB-CSIC), Campus UAB, E-08193 Bellaterra, Spain}
% Aurel: Mariano would prefer not being listed as a co-autor but to figure only
% in the Acknowledgement section.

\author{Alejandro R. Go\~ni}
%\email[Electronic mail:]{a.walsh@bath.ac.uk}
%\affiliation{Institut de Ci\`encia de Materials de Barcelona, 08193 Bellaterra, Spain}
\affiliation{ICREA, Passeig Llu\'is Companys 23, E-08010 Barcelona, Spain;
    Institut de Ci\`encia de Materials de Barcelona (ICMAB-CSIC), Campus UAB,
    E-08193 Bellaterra, Spain}

\author{Aur\'elien M. A. Leguy}
\author{Piers R. F. Barnes}
%\email[Electronic mail:]{a.walsh@bath.ac.uk}
\affiliation{Department of Physics, Imperial College London, SW7 2AZ, UK}

\author{Aron Walsh}
\email{a.walsh@bath.ac.uk}
\affiliation{Centre for Sustainable Chemical Technologies and Department of Chemistry, University of Bath, Claverton Down, Bath BA2 7AY, UK}
\affiliation{Global E3 Institute and Department of Materials Science and Engineering, Yonsei University, Seoul 120-749, Korea}
\date{\today}

\pacs{63.20.D−,63.20.Ry,78.30.-j}
%\keywords{Hybrid perovskite, phonons, Raman}

\begin{abstract}
%AW - one paragraph for easy pasting
%JMF - still an unsolved problem with git diffs :|
The hybrid halide perovskite \ce{CH3NH3PbI3} exhibits a complex structural behaviour, with successive transitions between orthorhombic, tetragonal and cubic polymorphs at ca. 165 K and 327 K. Herein we report first-principles lattice dynamics (phonon spectrum) for each phase of \ce{CH3NH3PbI3}. The equilibrium structures compare well to solutions of temperature-dependent powder neutron diffraction. By following the normal modes we calculate infrared and Raman intensities of the vibrations, and compare them to the measurement of a single crystal where the Raman laser is controlled to avoid degradation of the sample. Despite a clear separation in energy between low frequency modes associated with the inorganic \ce{(PbI3^-)_n} network and high-frequency modes of the organic \ce{CH3NH3+} cation, significant coupling between them is found, which emphasises the interplay between molecular orientation and the corner-sharing octahedral networks in the structural transformations. Soft modes are found at the boundary of the Brillouin zone of the cubic phase, consistent with displacive instabilities and anharmonicity involving tilting of the \ce{PbI6} octahedra around room temperature.
%\textbf{Keywords:} Hybrid perovskite, phonons, Raman
\end{abstract}

\maketitle

\section{INTRODUCTION}
Materials that adopt the perovskite crystal structure are known for their complex structural landscapes, with a large number of accessible polymorphs depending on the temperature, pressure, and/or applied electric field.
For ternary \ce{ABX3} perovskites, the A site cation is at the centre of a cube formed of corner sharing \ce{BX6} octahedra.
Displacement of the A cation is usually associated with a ferroelectric (Brillouin zone centre) instability,
while tilting of the \ce{BX6} octahedral network is usually linked to antiferroelectric (Brillouin zone boundary) transitions.\cite{Glazer1972,Benedek2013a}

Hybrid organic-inorganic perovskites are formed when one of the elemental perovskite building blocks is replaced by a molecular
anion or cation.\cite{Mitzi2001,Frost2014b}
There exists a large family of such compounds, 
including the widely studied formate perovskites,
which contain both molecular anions and cations.\cite{Cairns2013,Li2013e,Kieslich2015} 
Hybrid \textit{halide} perovskites are of current intense research effort,
due to their high efficiency photovoltaic action.\cite{Kojima2009,Im2011,Lee2012,Liu2013,DeAngelis2014,Even2015a,Walsh2014b,Walsh2015a,Jeon2015a}

Methylammonium lead iodide (\ce{MAPbI3}, where \ce{MA} represents the
\ce{CH3NH3+} cation), was first reported by Weber in 1978.\cite{Weber1978} 
It is the most relevant hybrid halide perovskite for photovoltaic application.
The transition from orthorhombic to tetragonal to cubic perovskite structures
as a function of temperature has been studied by techniques including
calorimetry and infrared spectroscopy\cite{Onoda-Yamamuro1990}, 
single-crystal X-ray diffraction\cite{Kawamura2002}, and dielectric spectroscopy.\cite{Onoda-yamamuro1992}
Recently analysis of powder neutron diffraction (PND) measurements has provided more quantitative insights into the temperature dependent behaviour of the MA cation within the anionic \ce{(PbI3^-)_n} network.\cite{Weller2015a}
There is now direct evidence for the degree of order of the MA cation in the
different phases, and the average lattice parameters (and thus extent of octahedral
tilting) as a function of temperature through the first and second order phase
transitions.
Quasi-elastic neutron scattering has provided further insights into the rotational dynamics of
the MA cation with a room temperature residence time of $\sim$ 14 ps,\cite{Leguy2015b}
while time-resolved vibrational spectroscopy identified fast librations (300 fs) 
and slow (3 ps) rotations of the molecule.\cite{Bakulin2015a} 

In this study, we calculate the phonon dispersion in each phase of \ce{MAPbI3}
within the harmonic approximation,
computing the force constants with density functional theory (DFT). 
We use the PBEsol functional, which is a generalised-gradient-approximation
(GGA) to the exchange-correlation functional, numerically evaluated with Perdew's
method, adjusted to give more accurate lattice constants and forces
for solids.\cite{Perdew2008a} 
The lattice dynamic calculations allow the atomic origin of each phonon mode
to be identified.
Changes in lattice polarisation and polarisability for each eigenvector
provide the infrared and Raman activity of each mode.
Spectral features related to the inorganic and organic components (from 0 to
3000 \icm) are well 
reproduced in comparison to the Raman spectra of a single crystal of \ce{MAPbI3}.
Overlap is found between the vibrations of the \ce{CH3NH3+} and \ce{PbI3-} components
up to 130 \icm, with the modes from 300 to 3000 \icm ~ being associated with pure molecular
vibrations.
The phonon dispersion has implications for developing quantitative models 
for the generation, transport and recombination of photo-generated electrons and holes
in hybrid perovskite solar cells. 

\subsection{Structure Models}
%
% Here we discuss the neutron diffraction structures in turn.
% 
The normal modes of a system are defined for an equilibrium
configuration. 
Calculating the vibrations for a non-equilibrium structure will result in
imaginary frequencies upon diagonalising the dynamical matrix. 
Therefore we have generated well
optimised structures of \mapi{}. 
One challenge in calculating the phonons of hybrid perovskites is the
 soft nature and complicated potential energy landscape of some of the
restoring potentials, particularly those involving the organic cation. 

The models for the crystal structures used in this study are discussed in detail below
and a comparison with the measured diffraction patterns are provided as Supplementary Material.\cite{si}

% JMF: TODO: nice potential energy landscape diagram? Mmmm...
% Selective dynamics with the MA frozen to sample some of the 6-deg freedom?

\begin{table*}[t]
%\begin{ruledtabular}
\begin{tabular}{lccccccccc}
Phase      & \textit{a} (\AA)    & \textit{b} (\AA)    & \textit{c} (\AA)    & $d$(Pb--I) (\AA) & \textit{Z} & Cut-off (eV) & \textit{k}-points	& Forces (meV/\AA) & $\Delta$H (meV)\\
\hline
\textit{Orthorhombic}  \\
DFT/PBEsol & 9.04 & 12.66& 8.35 & 3.18 & 4	& 700  & 5 $\times$ 4 $\times$ 5 & $1 $ & 0\\ 
PND (100 K) & 8.87 & 12.63 &  8.58 & 3.19 \\
\textit{Tetragonal} \\    
DFT/PBEsol & 8.70 & 8.72  & 12.83 & 3.19 & 4	& 800 & 5 $\times$ 5 $\times$ 3 & $1$ & 2\\
PND (180 K)    &  8.81 & 8.81 & 12.71 & 3.17 \\
\textit{Cubic}  \\
DFT/PBEsol       & 6.29 & 6.23  & 6.37 & 3.17 	& 1 & 700 & 6 $\times$ 6 $\times$ 6 & $1$ & 90 \\
PND (352 K)       & 6.32 & 6.32  & 6.32 & 3.16 \\
\end{tabular}
\caption{Equilibrium cell parameters from DFT/PBEsol energy
    minimisation, including the converged plane-wave cut-off, \textit{k}-point
    mesh and force threshold. \textit{Z} represents the number of formula units
    of \ce{CH3NH3PbI3} per cell.  
    The calculated difference in enthalpy ($\Delta$H) of each phase is given
    with respect to the ground-state orthorhombic configuration and per
    \ce{CH3NH3PbI3} unit.
% All equilibrium structures are available in an on-line repository.\cite{structures}
% TODO: JMF: Github / Zenodo DOI reference for these structs...
    Shown for comparison are the cell parameters and average Pb--I inter-atomic separations (\textit{d}) from powder neutron diffraction (PND).\cite{Weller2015a}
}
\label{tab:structure}
%\end{ruledtabular}
\end{table*}

\subsubsection{Orthorhombic Phase}
The orthorhombic perovskite structure is the low temperature ground state of
\ce{MAPbI3} and maintains its stability up to ca. \unit[165]{K}. \cite{Poglitsch1987c,Baikie2013,Weller2015a}   
A  comparison of the enthalpy from DFT calculations confirms this ordering in stability.  
The difference in enthalpy is small, just 2 meV per \mapi{} unit
compared to the most stable tetragonal phase, yet 90 meV compared to the
high-temperature cubic phase.
% JMF: I think this also makes sense in terms of the 1st. vs. 2nd order phase
% transition

% It is 165 K for Weller; 162 for \cite{Onoda-Yamamuro1990}.
Initial diffraction pattern solutions assigned the $Pna2_1$ space
group.\cite{Poglitsch1987c,Onoda-Yamamuro1990} 
Recent analysis of higher quality powder neutron
diffraction data reassigns it to $Pnma$ (a $D_{2h}$ point group).\cite{Weller2015a}
%and excluded the other candidate $Pmc2_1$\cite{Baikie2013} symmetry(???).
% JMF: I haven't read Baikie2013, but if that is true, we should put it in
% JMF: TODO: I think we should also explicitly discuss the differences in
%   space-group; extra operation? different point group?
The structure is a $\sqrt 2 a \times \sqrt 2 a \times 2 a$ supercell expansion of the simple
cubic perovskite lattice, i.e. following the lattice transformation matrix
\begin{equation}
\left( \begin{array}{ccc}
1 & -1 & 0 \\
1 & 1 & 0 \\
0 & 0 & 2 \end{array} \right).
\end{equation}

In the $Pna2_1$ phase, the \ce{PbI6} octahedra are distorted and tilt as $a^+b^-b^-$ in
Glazer notation\cite{Glazer1972} with respect to the orientation of the conventional cubic cell.
In this low-temperature phase, the four molecular cations in the unit cell are
static on the diagonals of the $ab$ planes pointing towards the undistorted facets of the cuboctahedral cavity.
Correspondingly, molecules belonging to different planes are anti-aligned with a head-tail motif. 
Such an antiferroelectric alignment is expected from consideration of the molecular dipole-dipole interaction.\cite{Frost2014}
 
In the low temperature orthorhombic phase the \ce{CH3NH3+} sublattice is fully ordered (a low entropy state).
The ordering may be sensitive to the material preparation and / or cooling rate into
this phase, i.e. the degree of quasi-thermal equilibrium. 
It is possible that different ordering might be frozen into the low temperature
phase by epitaxy or application of external force or electric fields.

\begin{figure}[h]
\centering 
\includegraphics[width=\columnwidth]{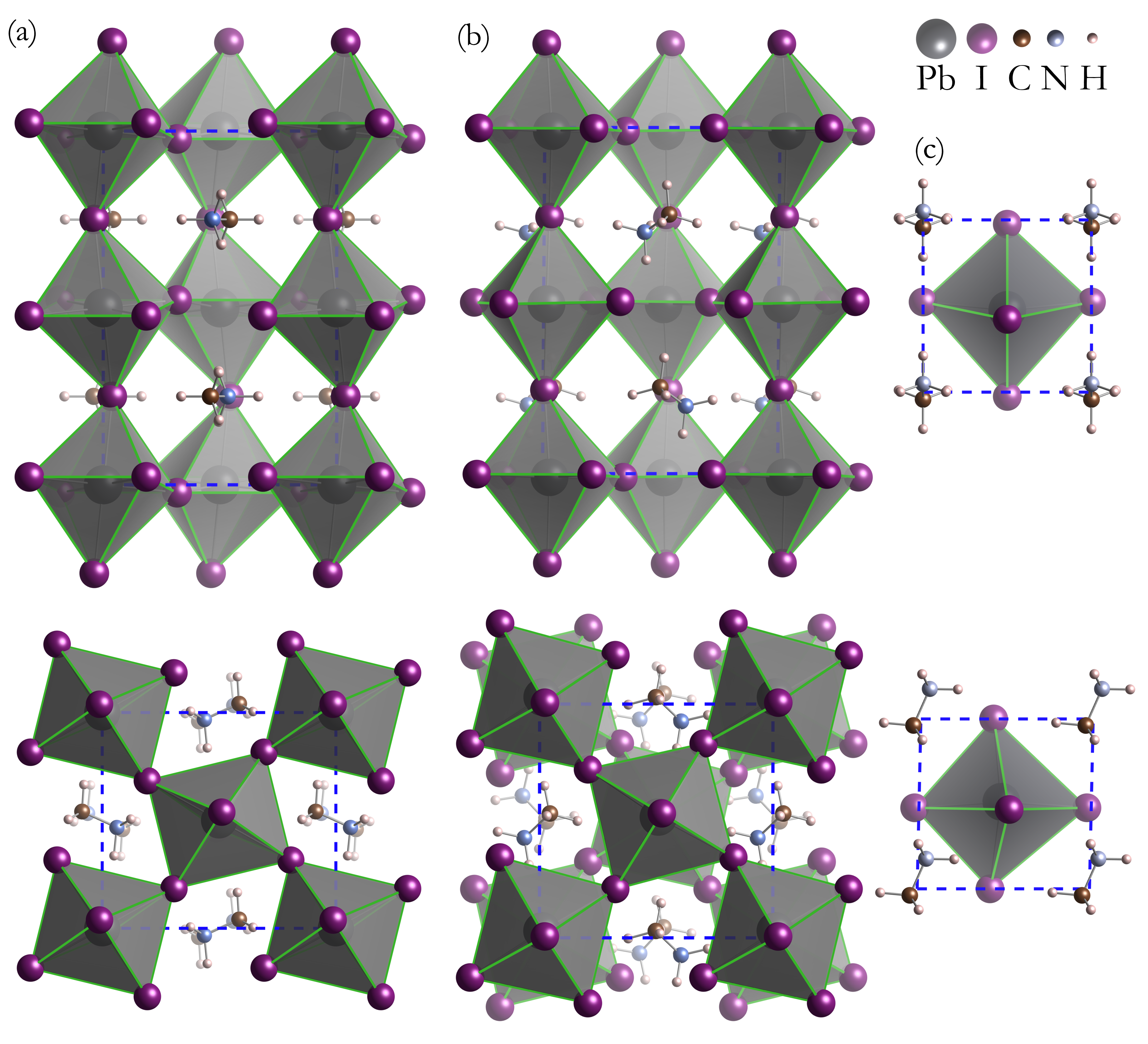}
\caption{
(Color online)
The crystal structures of the (a) orthorhombic, (b) tetragonal and (c) cubic phases of \ce{CH3NH3PbI3}. 
The upper and lower panels are oriented through $<100>$ and $<001>$, respectively.  
Lattice parameters and coordinates obtained from powder neutron diffraction were optimised using density functional theory (PBEsol).  
The \ce{PbI6} octahedra are shaded grey. 
All structures are available in an on-line repository.\cite{structures}
}
\label{fig:structures}
\end{figure}

\subsubsection{Tetragonal Phase}
At 165 K \mapi{} goes through a first-order phase transition 
from the orthorhombic to the tetragonal
space group $I4/mcm$ ($D_{4h}$ point group), which continuously undergoes
a second-order phase transition to the cubic phase by ca. 327 K\cite{Poglitsch1987c,Baikie2013,Weller2015a}.  
As with the orthorhombic phase, this can be considered a $\sqrt 2a \times \sqrt
2a \times 2a$ expansion of the cubic perovskite unit cell.  

The molecular cations are no longer in a fixed position as in the orthorhombic phase.
The molecules are disordered between two non-equivalent positions in each cage.\cite{Wasylishen1985,Baikie2013}. 
%from onoda-yamamuro not 4? 
The tetragonal distortion parameter in the cubic basis is greater than unity
($\frac{c}{2a} \sim 1.01$ at 300 K), corresponding
to an elongation of the \ce{PbI6} octahedra along the $c$ axis.
The associated octahedral tilting pattern is $a^0a^0c^-$ in Glazer notation.
% see http://link.aps.org/doi/10.1103/PhysRevB.89.125203

Atomistic simulations within periodic boundary conditions
 require an ordered configuration. 
The solved crystal structure shows that there are several possible
configurations for the organic cations within the tetragonal unit cell. 
These configurations have similar enthalpies within DFT,\cite{Fan2015a}
which is consistent with the observed disorder.
We choose to use the most energetically stable structure,
which is also consistent with a previous DFT investigation.\cite{Quarti2014a}

In the model of the  tetragonal structure, the MA cations are aligned as in the orthorhombic phase, towards the face of the perovskite cage, i.e. $<100>$ in the cubic basis.
The MA in different (001) planes are approximately orthogonal to one other.
The orientional dynamics of the methylammonium ions, which exists above 165 K, is not taken into account 
in this equilibrium configuration study.

\subsubsection{Cubic Phase}
With increasing temperature the tetragonal lattice parameters become more isotropic (i.e. $\frac{c}{2a}$ moves closer to 1), and the molecular disorder increases, to the point where a transition to a cubic phase occurs around 327 K.
The transition can be seen clearly from changes in the heat capacity,\cite{Onoda-Yamamuro1990} as well as in temperature dependent neutron diffraction\cite{Weller2015a}.

The cubic space group $Pm\bar{3}m$ ($O_h$ symmetry) has been assigned to this high-temperature phase. 
Although the methylammonium ions  posses $C_{3v}$ symmetry, the orientational disorder gives rise to the effective higher symmetry on average.
The local structure will necessarily have a lower symmetry. 
Indeed, for the bromide and chloride analogues of \mapi{}, pair-distribution function analysis of X-ray scattering data indicates a local structure with significant distortion of the lead halide framework at room temperature.\cite{Worhatch2008}

We previously considered alignment of the molecules along the principal $<100>$
(face), $<110>$ (edge), and $<111>$ (diagonal) directions of the cubic unit
cell, and showed that they are of similar DFT enthalpy, with a small barrier for rotation.\cite{Brivio2013}  
Further \textit{ab-initio} molecular dynamics showed an average preference for the
$<100>$ facial configuration at 300 K.\cite{Frost2014}
Therefore we chose the $<100>$ configuration as our reference structure for the lattice vibrations. 

Representations of the crystal structure of each phase are shown in Figure
\ref{fig:structures}, the equilibrium structure parameters are listed in
Table \ref{tab:structure}, and the structures themselves are available in an on-line repository.\cite{structures}

\begin{figure*}
  \includegraphics[width=\textwidth]{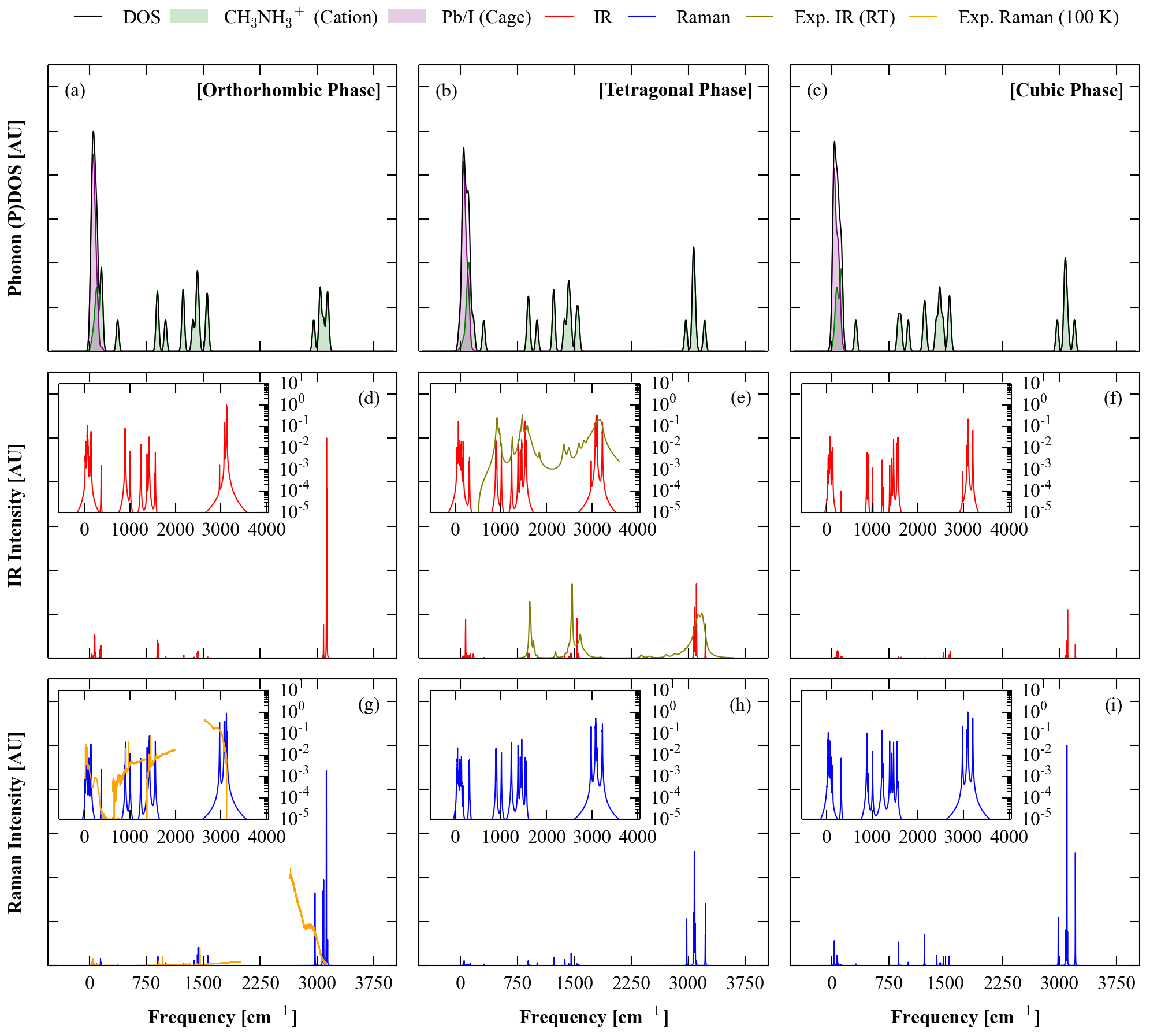}
  \caption{
  (Color online)
  (a-c) Projected phonon density of states (PDOS) for the three phases of \ce{CH3NH3PbI3}
   as calculated from DFT/PBEsol, generated by convolution with a 32 \icm{} Lorentzian.
   Note that no imaginary (negative) modes are found in the orthorhombic or tetragonal phases.
  (d-f) Simulated infrared (IR) spectra.
  (e) The measured IR spectrum at 300 K from Ref. \onlinecite{Glaser2015a} is coplotted in mustard (light grey).
  (g-i) Simulated Raman spectrum.
  (g) The measured Raman spectrum of a single crystal in the orthorhombic
phase at 100 K is coplotted in orange (light grey).
Simulated spectra were broadened by convolution of a 2 \icm{} Lorentzian. 
  Insets in the simulated spectra are the same data on a logarithmic scale to
  show the structure in the low intensity modes.
  }
\label{fig:phonons}
\end{figure*}
%\todoin{JMF: I think this should have some short comment on the spectral broadening methods used}
% NB: 10 \icm is what Jonathan used in his Python
% For DoS, Phonopy auto selects a value based on the width of the Phonon
% spectrum, this is listed at the top of the '*DOS.dat' datafiles (in THz).
% I took the average of the Tetra, Ortho and Cubic: .963956 THz
% Which converts to 32.15412 \icm, which I rounded to 32.

\section{METHODS}

\subsection{Computational}
The total energy and atomic forces were computed from first-principles within density functional theory as implemented in the code VASP.\cite{Kresse1996,Kresse1999}
Noise in the lattice vibrations was minimised by rigorous convergence of total
Kohn-Sham energy with respect to the basis set (kinetic energy cut-off for
plane waves) and sampling of reciprocal space (density of the \textit{k}-point
mesh). 
The final computational set-up is summarised in Table \ref{tab:structure}.

We performed complete optimisation of the cell volume, shape and atomic positions, with the PBEsol\cite{Perdew2008a} semi-local exchange-correlation functional.
The scalar-relativistic projector-augmented wave method\cite{Blochl1994} was
employed, with a pseudo-potential treating the Pb 5d orbitals as valence. 
Spin-orbit
coupling was not considered as it mainly affects the Pb 6p conduction 
band, which does not influence the interatomic interactions at equilibrium.
All atomic forces were reduced to below a threshold of 1 meV/\AA.
Due to the presence of the organic cations, which breaks the ideal lattice
symmetry, deviations in the expected parameters can occur, e.g. in the high
temperature pseudo-cubic phase, the three equilibrium lattice parameters are
not equal.  
The equilibrium structure parameters (at 0 K and excluding zero-point
contributions) are reported in Table \ref{tab:structure}.

The normal modes are calculated within the harmonic approximation, using the
Phonopy\cite{Togo2010,Skelton2014,Togo2015} package to construct and evaluate
the dynamical matrix composed of DFT force constants. 
Both the finite displacement method (FDM or supercell approach)\cite{Stoffel2010} 
and density functional perturbation theory (DFPT)\cite{Baroni2001} approaches
to construct the force constants were tested.
The results of both approaches produced similar vibrational 
spectra, 
with a variance in the mode energies of 6 cm$^{-1}$.

Within a primitive cell of \textit{N} atoms there are 6\textit{N} possible
displacements ($\pm x,\pm y,\pm z$), which can be reduced by the crystal
symmetry.  
For the orthorhomic phase, the 288 possible displacements are reduced to 41, while
the tetragonal and cubic phases required 288 and 72 displacements, respectively. 
The phonon dispersion (for \textbf{q}-points away from the Brillouin zone
centre, the $\Gamma$ point) in the cubic phase was probed in a $2\times2\times2$ supercell. 
Due to computational expense, we do not calculate this for the other (larger
unit cell) phases.
For the phonon density of states, Brillouin-zone integrations were performed with 
$36\times36\times36$ (orthorhombic and tetragonal) and $48\times48\times48$  
(cubic) $\Gamma$-centered Monkhorst-Pack \textit{q}-meshes.

Once the normal eigenmodes and eigenvalues are calculated, it is
possible to model their associated Raman and infrared (IR) activity by mode
following. 
The two spectroscopic techniques probe different physical responses of the material: 
the change in polarisation for IR and the change in polarisability for Raman.
The IR spectra are simulated with the analytic formula of Gianozzi \& Baroni
(using the Born effective charge tensor)\cite{Baroni2001}.
Prediction of the Raman spectra required computing the change in macroscopic
dielectric tensor with respect to each normal mode of the system, a significant
DFT calculation in terms of computational expense.\cite{Skelton2014b}

\subsection{Experimental}
Methylammonium lead iodide single crystals were grown according to the method of Poglitsch and Weber.\cite{Poglitsch1987c}
12.5 g of lead acetate trihydrate (\ce{Pb(CH3CO2)2.3H2O}, Sigma) was dissolved in 10 mL hydroiodic acid (HI$_{aq}$, 57 wt\%, Sigma) in a 50 mL round bottom flask and heated to 100$^{\circ}$C in an oil bath. 
Separately, 0.597 g of \ce{CH3NH2} (aq, 40 wt\%, Sigma) was added dropwise to
a further 2 mL of HI$_{aq}$ kept at 0$^{\circ}$C in an ice bath under
stirring.  
The methylammonium iodide solution was then added to the lead acetate solution
and the mixture was cooled over five days to a temperature of 46$^{\circ}$C,
resulting in the formation of black crystals with largest face length around
8 mm.  
The content of the flasks was subsequently filtered and dried for 12 hours at
100$^{\circ}$C.

Raman spectra were collected in backscattering geometry with a high resolution
LabRam HR800 spectrometer using a grating with 600 lines per millimetre and
equipped with a liquid-nitrogen-cooled charge coupled device (CCD) detector.
The 785 nm line of a diode-pumped solid state laser was used as excitation beam
and focused onto the sample using a long distance 20$\times$ microscope
objective. 
Raman
measurements were carried out at 100 K using a gas-flow-type
cryostat with optical access that fits under the microscope of
the Raman setup. 
%% NEW
The high spectral resolution and stray-light
rejection of the LabRam spectrometer, particularly in
combination with the 785 nm line, allowed us to measure the
Raman spectrum of MAPI at very low Raman shifts down to 20 to
30 cm$^{-1}$. This way and for the first time, we were able to
spectrally resolve several low-frequency modes associated to
vibrations of the inorganic cage of the hybrid perovskite.

Heating by laser light directly absorbed by \ce{CH3NH3PbI3} has been shown to lead
to rapid degradation of the material resulting in \ce{PbI2} Raman
signatures.\citep{ledinsky2015}
Since 785 nm light is only weakly absorbed the heating effect of the laser  was
low enough to ensure the crystal structure was preserved. The power density
incident on the sample was kept at ~80 W/cm$^2$. 
At such power level, it was checked that no
appreciable spectral changes in peak width and/or position
occurred, yet maintaining a good signal-to-noise ratio.
Further, samples were kept under vacuum inside the cryostat
during the measurements.

\section{RESULTS}

\subsection{Harmonic Phonons}
The full phonon density of states (DOS) is shown for the three phases of
\ce{CH3NH3PbI3} in Figure \ref{fig:phonons}. 
Also plotted is the partial DOS, where assignment to \ce{CH3NH3+} or \ce{PbI3-}
is performed based on the atomic contribution to each eigenvector.
An animation of all 36 eigenmodes of the cubic phase is provided as Supplementary Material\cite{si}.

Qualitatively, each \mapi{} phase shows similar vibrational properties with
three energetic regions of phonons: 
(\textbf{i}) a low-frequency band from 0--150 \icm;
(\textbf{ii}) a mid-frequency band from 280--1600 \icm;
(\textbf{iii}) a high-frequency band from 2900--3300 \icm.
These ranges are consistent with previous computational reports.\cite{Quarti2013,Ahmed2014}

Due to the large difference in atomic mass of the organic and inorganic
components, and to the difference in bonding between the inorganic cage and the
covalently bonded molecule, 
we anticipated that the low frequency modes would comprise entirely of motion
by the \ce{PbI6} octahedra, while the high frequency modes will involve the \ce{CH3NH3+} cation.
This is qualitatively the case, but there is also significant coupling between the two. 
Taking the example of the cubic phase with 36 modes:
the highest-energy 18 modes (forming bands \textbf{ii} and \textbf{iii}) correspond to molecular vibrations, i.e. the 3\textit{N}-6
modes of the methylammonium ion. 
For an isolated non-linear molecule, the 6 translational and rotational degrees
of freedom do not contribute to the pure vibrational spectrum, but this is not
the case for a molecule inside a cuboctahedral cavity.

The 6 additional molecular modes are strongly coupled to the 9 modes (i.e. 3$N$-3 for \ce{PbI3-}) associated with stretching of the Pb--I bonds and breathing of the \ce{PbI6} octahedra, which results in the spectral overlap observed in the partial DOS of band \textbf{i}. 
Particularly striking is the low-frequency pivoting motion associated with the libration of the molecular dipole, coupled with a breathing of the octahedral framework (e.g. modes 10 and 15 in the SI).
The final three zero-frequency modes correspond to acoustic translations of the lattice.

%\begin{figure}
%  \includegraphics[width=\columnwidth]{./img/vibs.png}
%  \caption{
%  Static representation of the two $\Gamma$-point phonon modes in the cubic
%  phase of \mapi{}. 
%  The black arrows represent the direction and magnitude of the eigenvector on each atom.  
%  The left and right panels show modes 10 (58 \icm) and 15 (92 \icm), respectively.  An animation of all 36 modes is provided as SI.
%  [JMF: Intend to replace this with two molecular eigenmodes...]
%  }
%\label{fig:vibs}
%\end{figure}

\begin{figure*}
  \includegraphics[width=\textwidth]{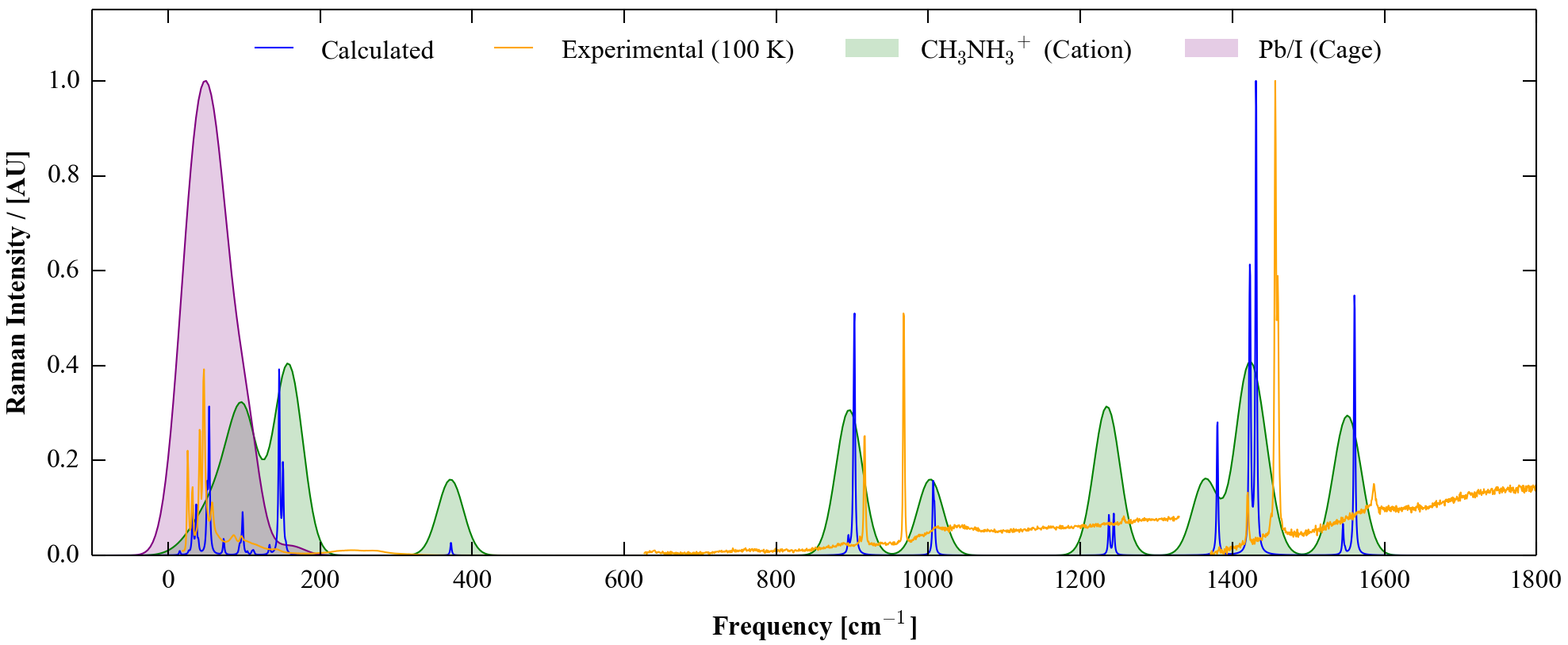}
  \caption{
  (Color online)
 Comparison of the calculated and measured (100 K) Raman spectrum for the orthorhombic perovskite phase of \ce{CH3NH3PbI3}
 in the range 0--1800 \icm{}, which includes phonon modes centred on the inorganic cage (lowest frequency), organic cage (highest frequency) and coupled-modes (intermediate frequency). 
 The simulated spectrum was broadened by convolving a 2 \icm{} Lorentzian (to
 match experimental broadening of the high resolution Raman spectrometer),
 while the underlying phonon density of states is shown for comparison with
 a broadening of 0.5 THz.
  }
\label{fig:raman}
\end{figure*}

\subsection{Vibrational Spectra}
For \mapi{} it is not possible for the relaxed (non-idealised) crystal structures
to assign the spectral activity directly 
with
group theoretic irreducible representation analysis of the phonon modes.
The molecule breaks the average crystal symmetry. 
This symmetry lowering allows for simultaneous Raman and IR activity even in the pseudo-cubic phase. 
The predicted spectra (the $\Gamma$ point phonon modes weighted by the computed
spectral intensity, convolved with a Lorentizian for experimental comparison) are reported in Figure \ref{fig:phonons} for each phase.

Raman and IR activity is observed across each of the three phonon bands
previously discussed. 
A notable exception is the lowest-energy purely molecular vibration near 300 \icm, 
which is neither Raman nor IR active.

To understand the effect of embedding the methylammonium in \mapi{}, we calculate
the normal mode vibrations of an isolated methylammonium ion in vacuum with
a similar density functional theory method to our periodic calculations 
(PBE with an atom centered augmented cc-pVQZ basis set). 
Thereby we calculate the 18 molecular modes, directly accessing their symmetries
and nature. 
The C$_{3v}$ symmetry of the molecule separates the vibrational bands into one
$A$ symmetric mode and blue-shifted two-fold degenerate
asymmetric $E$ modes.
The six bands we find are in ascending energy: 
twist around the C--N axis (282, 886 \icm);
vibration along the C--N axis (923, 1239 \icm);
bending of the C--H bonds (1418, 1451 \icm);
bending of the N--H bonds (1478, 1622 \icm);
stretching of the C--H bonds (3018, 3119 \icm);
stretching of the N--H bonds (3321, 3395 \icm).

As the cation charge density is centred towards the N, the
motion of the protons associated with N have the strongest affect on the dipole moment
and therefore strongest IR activity. 
Due to the stronger bonds, their frequencies are
consistently blue shifted relative to the C end.
Owing to the molecular dipole moment 
(2.2 D,\cite{Frost2014} which is rotation and position invariant, 
and corresponds to a polarisation contribution of $\sim$ 3 $\mu$C/cm$^{2}$), the
two high-frequency asymmetric stretching modes of \ce{NH3+} (band \textbf{iii}) 
results in the strongest absolute IR intensity. 
The hydrogen stretching modes (band \textbf{iii}) are responsible for significant Raman activity.
The only mode involving C or N motion is the weakly IR and Raman
active vibration at 923 \icm (vacuum), 1007 \icm (cubic perovskite).

The rotation of the \ce{CH3} against the \ce{NH3} unit, while being strongly
populated in molecular dynamic simulations, and which forms the main
source of quasi-inelastic neutron scattering, 
%\cite{Leguy2015}, 
is entirely IR and Raman inactive in vacuum. 
This is the mode responsible for the 282 \icm (vacuum), 318 \icm (cubic)
, 300--310 \icm (tetragonal), 370--372 \icm (orthorhombic) vibration. 
Progressive confinement of MA from vacuum to the orthorhombic phase
blue-shifts the energy of the vibration.

In the solid state, the degeneracies in the molecular modes are
typically split by local environment effects, peaks are both redshifted and
blue-shifted, and the IR and Raman activity varies. 
As such, it is evident that analysis of the Raman and IR spectra in the
experimentally easily accessible molecular frequency range can enable
statements to be made about the local structure and configuration of the hybrid
perovskite. 
Our data is collected for a particular representation of the cubic and
tetragonal phase; in reality the location of the MA in these
phases will be disordered. 
As such, detailed comparison of theory to experiment will require 
sampling the thermodynamic ensemble of structures. 

To our knowledge, no Raman spectra for the three phases have previously been reported across the full frequency range. 
The spectrum up to 450 \icm ~ was reported in Ref. \onlinecite{Quarti2013}.
Reliable measurements are a challenge due to chemical instability of the
material.
\ce{MAPbI3} is strongly affected by environmental conditions, such as the
presence of ambient moisture\citep{Baikie2013,Noh2013a}.
Isolated in vacuum the material can still decompose and bleach due to
heating, including by that imposed by the (typically high) Raman laser fluence.\citep{ledinsky2015}
Such degradation leads to the formation of \ce{PbI2}, which overlaps in Raman
spectra with \mapi{}, and so easily leads to misinterpretation. 

The Raman spectra of a high-quality single crystal of \mapi{} is shown in Figure \ref{fig:phonons},
and compared in detail with the calculations in Figure \ref{fig:raman}.
Across the full spectral range, 
the agreement between the predicted and measured spectra is good, 
with the response across bands \textbf{i}, \textbf{ii} and \textbf{iii}  well reproduced.
On closer inspection in Figure \ref{fig:raman}, there are noticeable shifts in peak positions,
which can be attributed to three potential sources of error:
(a) the harmonic approximation (anharmonic renormalisation may be large);
(b) the limits of the exchange-correlation treatment (non-local interactions may be important);
(c) the assumption of a fully-ordered structure (local inhomogenity may be prevalent).
There is also a notable case of a missing peak around 150 \icm,
which we can tentatively attribute to a lifetime broadening effect.
The same level of theory applied to the lead based semiconductors PbS and PbTe results in 
quantitative agreement with measured phonon frequencies and dispersion,\cite{Skelton2015z}
which highlights the complex nature of \ce{CH3NH3PbI3}.

We have also included in Figure \ref{fig:phonons} a room temperature IR spectrum reported
by Glaser et al.\cite{Glaser2015a},
which again shows excellent agreement across the spectrum. 
A number 
of very weak absorption peaks below 3000 \icm ~ are evident, which could be 
related to molecular disorder and/or partial decomposition. 
The temperature resolved (between 140 and 299 K) IR spectra for
tetragonal and orthorhombic phases have been previously reported by Yamamuro\citep{1992}. 
We reproduce the position and the intensity of the peak observed at 900 \icm,
reliably assigning it to the (also Raman active) C--N bond stretch. 
High quality IR measurements, in particular looking at the very low energy
transitions, would provide considerable information on the nature of the
domains and local structure in a \mapi ~ film. 

%In literature are reported the following peaks for Raman
%park2015 71, 94,97, 108,110
%quarti2013 62, 94, 119, 154, 250,390
%Quarti2013 DFT with LDA 
%safdari2014 molecule at 1500-1600, and NH at 2900-3000
%ledinsky15 52, 110 (tetragomnal phase)

\subsection{Anharmonic Effects}
The lattice dynamic simulations discussed above were performed within the harmonic approximation. 
All eigenmodes at the centre of the Brillouin zone were real (positive frequencies) for each phase, \textit{i.e.} the structures are locally stable. 

\begin{figure}[t!]
  \includegraphics[width=\columnwidth]{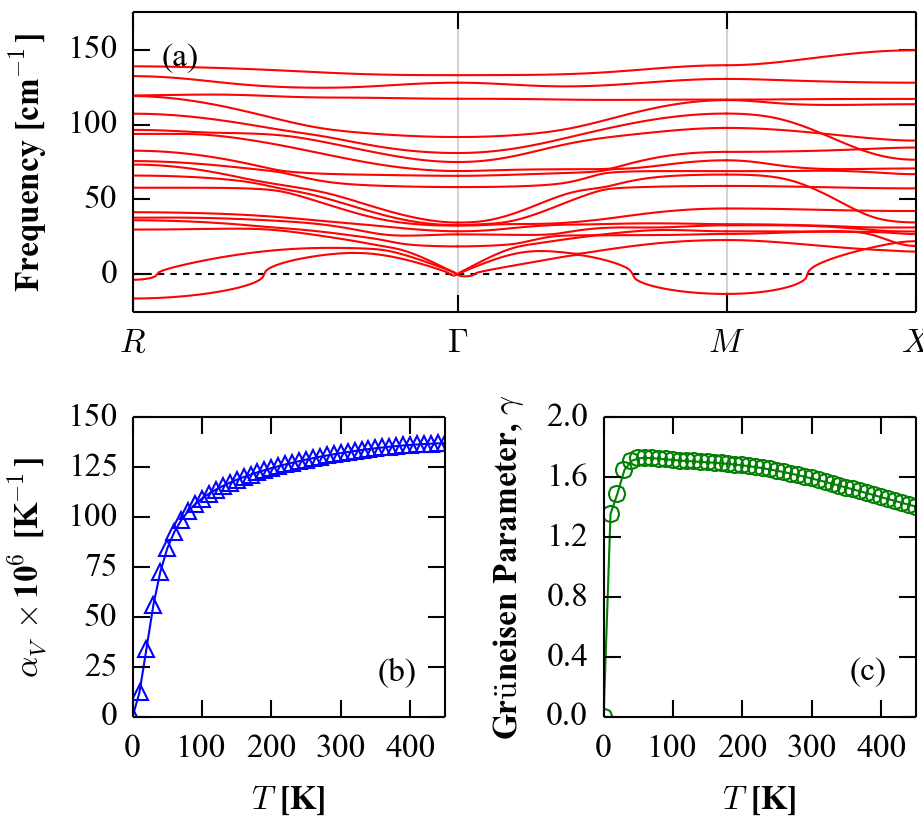}
  \caption{
  (Color online)
      Vibrational properties of the cubic phase of \mapi{}:  
   (a) Phonon dispersion of the low frequency (inorganic cage) 
   modes within the harmonic approximation. 
   Negative frequency 
   (imaginary or `soft') modes are found at the Brillouin zone boundary points
   $M ~(\frac{2 \pi}{a}(\frac{1}{2},\frac{1}{2}, 0))$ and 
   $R ~(\frac{2 \pi}{a}(\frac{1}{2},\frac{1}{2},\frac{1}{2}))$.
  (b) Volumetric thermal expansion within the quasi-harmonic approximation.
  (c) Average \grun ~ parameter within the quasi-harmonic approximation.}
  \label{fig:qha}
\end{figure}

The phonon dispersion across the first Brillouin zone is shown for the cubic perovskite structure in Figure \ref{fig:qha}.
Here imaginary (negative frequency or `soft') modes are found at the zone boundaries.
Such instabilities are a common feature of the perovskite structure, and 
represent antiferroelectric distortions linked to to rotations and tiling of the octahedra in neighbouring unit cells.\cite{Benedek2013a} 
The soft modes are centred around the \textit{R} and \textit{M} points, which correspond to the $<111>$
and $<110>$ directions in the cubic lattice.
This behaviour is similar to the inorganic perovskite \ce{CsPbCl3}, where neutron scattering
was used to probe condensation of these modes, which leads to successive 
transitions from the cubic to tetragonal to orthorhombic phases.\cite{Fujii1974}
The effect of these modes in \mapi{}, and the associated high levels of anharmonicity at room temperature 
can be observed directly in molecular dynamics simulations, where temporal rotations of the \ce{CH3NH3+} ions and distortions of the \ce{PbI6} octahedra have been found in several studies.\cite{Leguy2015b,Frost2014,Quarti2015}

An approach to including the effects of temperature (thermal expansion) and first-order anharmonicity in lattice dynamic calculations is the quasi-harmonic approximation (QHA).\cite{Dove1997a,Buckeridge2013b}
The computational cost is one order of magnitude higher than the harmonic approximation and thus was considered for the cubic phase only.
The volumetric thermal expansion coefficient extracted from the PND data at 300 K is $1.32\times10^{-4}/$K, which 
compares very well to the value of $1.25\times10^{-4}/$K computed within the QHA.
The predicted thermal expansion for \mapi{} is similar to inorganic semiconductors (e.g. for PbTe the value is $0.7\times10^{-4}/$K at 300 K\cite{Skelton2014}) and positive over the full temperature range. 

The temperature dependence of the phonon modes can be described by the \grun ~ parameter, which has an average of around 1.6 (see Figure \ref{fig:qha}), slightly below the value of 1.7 found in \ce{PbI2}\citep{Sears1979}. 
The imaginary modes at \textit{R} and \textit{M} remain at all temperatures,
 consistent with the cubic lattice being a dynamic average of a locally distorted structure;
 the same phenomenon is observed in \ce{CsSnI3}.\cite{Silva2015}
The high level of anharmonicity associated with the soft titling modes is consistent with the 
`ultra low' ($<$ 1 Wm$^{-1}$K$^{-1}$ at 300 K) lattice thermal conductivity reported for single crystals and polycrystalline \mapi.\cite{Pisoni2014} 
Hybrid halide perovskites are thus also promising for application in thermoelectric devices
if thermal stability issues can be overcome.\cite{He2014}

\section{CONCLUSIONS}
The vibrational frequencies of three crystallographic phases of the hybrid perovskite \ce{CH3NH3PbI3} have been investigated. 
We identified three main phonon branches present in the three phases.
Two high-frequency branches are associated with the vibration and bond
stretching of the molecular cation with frequencies in the range 300 \icm ~ to
3300 \icm.
The lowest energy branch, below 150 \icm, arises predominately from the inorganic cage,
but with half the modes coupled to the motion of the molecule. 
The simulated Raman spectrum is in good agreement 
with 
measurements on a single crystal of \mapi.
Dynamic instabilities occur at the zone boundaries, which requires methods 
beyond the harmonic approximation, such as self-consistent phonon theory, for an accurate treatment. 
These results suggest that the room temperature structure of \mapi ~ is 
fluctional, owing to the persistent titling and distortion of 
the octahedral networks and rotations of the molecular cations.

These factors may be important for developing a quantitative understanding 
and model
of how hybrid perovskite solar cells operate.
Upon excitation, the relative stability of free carriers and excitons depends
intimately on
the dielectric screening of the material, which includes vibrational and rotational components.
The transport and recombination of photo-generated 
charge carriers will also be influenced by electron-phonon coupling, 
which can significantly reduce the effective size and distribution of electrons and holes
within the perovskite layer.

\section{Web Enhanced}
 An animation of the 36 $\Gamma$-point phonon modes of the cubic phase of \mapi{} in gif format is available at \url{http://people.bath.ac.uk/aw558/temp/mapi_phonon.gif}.

\begin{acknowledgements} 
The authors are grateful for helpful discussions with Mariano Campoy-Quiles. 
The research at Bath has been supported by the EPSRC (Grant No. EP/K016288/1 and  EP/M009580/1), the ERC (Grant No. 277757), EU-FP7 (Grant No. 316494), and the Royal Society. 
AJJ and OJW were funded through the CDT in Sustainable Chemical Technologies (EPSRC Grant No. EP/ G03768X/1).
% Following from Piers:
PRFB and AMAL are grateful to the EPSRC (Grant Nos. EP/J002305/1, EP/M014797/1,
and EP/M023532/1). 
ARG acknowledges the Spanish Ministerio de Economía y Competitividad (MINECO) through project number MAT2012-37776 and CSD2010-00044 (Consolider NANOTHERM).
  This work benefited from access to both the University of Bath's High Performance Computing Facility and ARCHER, the UK's national high-performance computing service, which is funded by the Office of Science and Technology through EPSRC's High End Computing Programme (Grant No. EP/L000202). 
\end{acknowledgements}

\bibliography{library,library_extra}

\end{document}